\documentstyle[epsf,twocolumn,prl,aps]{revtex}

\newcommand{\be}{\begin{equation}}
\newcommand{\ee}{\end{equation}}
\newcommand{\bea}{\begin{eqnarray}}
\newcommand{\eea}{\end{eqnarray}}
\newcommand{\lton}{\mathrel{\lower.9ex
                  \hbox{$\stackrel{\displaystyle <}{\sim}$}}}

\begin{document}                                                
\title{Event-by-Event Fluctuations from Decay of a Polyakov Loop Condensate}
\author{Adrian Dumitru$^a$ and Robert D. Pisarski$^b$}
\bigskip
\address{
a) Department of Physics, Columbia University, New York, New York 10027, USA\\
email: dumitru@nt3.phys.columbia.edu\\
b) Department of Physics, Brookhaven National Laboratory,
Upton, New York 11973-5000, USA\\
email: pisarski@bnl.gov\\
}
\date{\today}
\maketitle
\begin{abstract} 
A model for particle production at the deconfining phase transition
in QCD is developed, as the semi-classical decay of a 
condensate for the
Polyakov loop.  In such a model, generically particle
production, as measured on an event-by-event basis, exhibits significant
deviations from statistical behavior.
\end{abstract}
\pacs{}

The deconfined phase of QCD at high temperature may be produced in
the collisions of large nuclei at high energies, such as at RHIC and LHC.
The usual picture of the deconfined phase is of a weakly interacting
gas of quasi-particles.  While this picture applies (well) above
the critical temperature, $T_c$, it may not be very useful near
$T_c$.  Recently, an effective theory which is better suited near
$T_c$ was proposed; in a mean field approximation, the free energy is
due to a potential for the Polyakov loop \cite{rp}.

In this Letter we propose that if the mean field theory is correct,
particle production at the phase transition is driven by coherent 
oscillations of the Polyakov loop condensate.  This is very similar
to particle production from preheating in inflationary models of 
the early universe \cite{preheating}.  
For both inflation and the mean field theory of \cite{rp}, 
the potential energy of the Polyakov loop {\it dominates} the total
energy density, so that particle 
production can be estimated by semi-classical means.  
Particle production from a coherent source is observable by
measuring fluctuations on an event-by-event basis, and differs
from particle production by an incoherent, or statistical, source.

Many of the features of our model
are similar to particle production in Disoriented
Chiral Condensates (DCC's) \cite{dcc1,dcc2}.  
Because pions are light, the chiral condensate 
in a DCC has a small energy density relative to the total energy density. 
Thus the number of produced particles from the DCC is small relative
to the total, and the particles are produced in a narrow region of phase space,
on top of a large, incoherent background.
In contrast, in the present model essentially {\it all}
of the energy of the deconfined phase is going into oscillations of
the Polyakov loop, and thereby into pions.  
We stress that while the Polyakov loop is treated classically, the
production of pions is quantum-mechanical.

We also contrast this with the conjectured critical end-point of 
the chiral phase transition in the plane of temperature, $T$, and
(baryon) chemical potential, $\mu$~\cite{jackson,rss}.
Through critical fluctuations in the sigma field, a critical end-point
will also generate non-statistical fluctuations.  The critical end-point,
however, only exists for a specific value of $T$ and $\mu$;
it is necessary to "tune" the collision parameters to reach this 
exceptional point.
If the effective theory of~\cite{rp} applies to QCD, however,
it should apply, about the critical temperature, for a large range of
chemical potentials near zero; there is no need for tuning.  
Further, critical fluctuations~\cite{jackson,rss} are
necessarily largest about zero momentum.  
In contrast, in the present model the produced particles 
typically have momenta of order several
hundreds of MeV, with significant fluctuations about
that large momentum scale.  Unlike a critical end-point, one
should not cut on particles with low momentum; deviations
from statistical behavior appear in the average pion momentum.

We begin by reviewing the mean field theory for the Polyakov loop~\cite{rp}, 
and then present an illustrative calculation of particle
production.  This calculation is not meant to be definitive,
but merely to sketch how
a classical condensate for the Polyakov loop might decay.

To understand the mean field theory, consider first 
an $SU(N)$ gauge theory not for three
colors, but in the limit of a large number of
colors, $N \rightarrow \infty$.  As first observed by Thorn \cite{thorn}, 
in this case the free energy itself is an order parameter.  For
temperatures below the deconfining phase transition, confinement
occurs, and the free energy is due exclusively to hadrons,
such as glueballs.  As color singlets, glueballs only contribute
of order one to the free energy.  We assume the usual, ``quarkless''
large $N$ limit, in which the number of
light quark flavors, $N_f$, is held fixed as $N \rightarrow \infty$;
then mesons also contribute of order one to the free energy.

In contrast, above $T_c$ deconfinement occurs, and 
gluons contribute $\sim N^2$ to the free energy;
the quark contribution is suppressed, $\sim N^2 (N_f/N)$.  
At very high temperature, 
by asymptotic freedom the behavior of the free energy can be computed
perturbatively.

There is a puzzle, however.  The free energy is a gauge invariant
quantity, and so at any temperature, 
should be describable {\it exclusively}
in terms of gauge invariant excitations.  While the glueballs change
their nature with temperature, they remain the dominant gauge invariant
excitations.  But if color singlet glueballs can only contribute
of order one to the free energy, 
what is the term $\sim N^2$ in the free energy due to?  

The only quantity which can provide such a contribution is an 
expectation value for the Polyakov loop:
\begin{equation}
\ell(x) = \frac{1}{N} \;
{\rm tr} \left( {\cal P} \exp\left( i g \int^{1/T}_0 A_0(x,\tau) \, 
d\tau \right)\right) \; ;
\label{e1}
\end{equation}
$\rm tr$ is the matrix trace, 
$\cal P$ is path ordering, and $g$ is the gauge coupling constant;
$A_0(x,\tau)$ is the time component of the vector potential
in the fundamental representation, at a spatial position $x$ and
euclidean time $\tau$.  At a temperature $T$, the Wilson line is
defined in imaginary time; we assume that it can be generalized to
real time, although its detailed form is as yet unknown \cite{rp}.
The full thermal Wilson line is an $SU(N)$ matrix; its expectation value
includes both color adjoint degrees of freedom, $\tilde{\ell}_a$,
and a color singlet, $\ell$ \cite{rp}.  The $\tilde{\ell}_a$'s
transform under the $SU(N)$ gauge group,
while the $\ell$'s only transform under a global $Z(N)$
symmetry.  The $SU(N)$ Wilson lines, $\tilde{\ell}_a$, may be important in
driving the transition first order when $N \geq 3$, but 
in a mean field approximation for the free energy,
only the Polyakov loop, $\ell$, matter \cite{rp}.
(Following standard convention, but contrary to \cite{rp}, henceforth
we refer to the $\ell$'s as $Z(3)$ Polyakov loops, and to the 
$\tilde{\ell}_a$'s as $SU(3)$ Wilson lines.)

As the Polyakov loop is a color singlet, its expectation
value, $\ell_0 = \langle \ell \rangle$, is gauge invariant.
This vanishes in the confined phase, up to
corrections $\sim N_f/N \sim 1/N$, and 
is nonzero above $T_c$.  At infinite temperature, by asymptotic
freedom we can ignore fluctuations in the gauge field, and
$\ell_0 \rightarrow 1$.

Thus the term $\sim N^2$ in the free energy is due
exclusively to the potential for the Polyakov loop condensate,
${\cal V}(\ell)$.
The free energy of glueballs 
and other gauge invariant excitations contribute $\sim 1$,
while quarks contribute $\sim N^2 (N_f/N) \sim N$.

Of course $N$ is not infinite in QCD, but three.  
At least for the free energy, perhaps large $N$ is a good guide to
the behavior for $N=3$.  If so, then in the high temperature phase
of QCD, the free energy is --- in a precise sense ---
dominated by gluons, through the potential for the Polyakov loop.
This is supported by Lattice data, which
finds that the free energy is much smaller at low,
than at high, temperature \cite{rp}.

There is one aspect of the deconfining phase transition for $N=3$
which is very different from $N=\infty$.  General arguments and Lattice
simulations \cite{four} suggest that the deconfining transition is of first
order when $N \geq 4$ (it may be strongly first order, although 
definitive results in the continuum limit are lacking).  In contrast,
the deconfining phase transition is of second order for two colors \cite{two},
and {\it nearly} second order for three colors \cite{three}.  
Thus at least as far as the order of the deconfining transition in the pure
glue theory is concerned, $N=3$ is ``near'' $N=2$, not $N=\infty$.  

For $N=3$, the order of the transition does change
with the addition of dynamical quarks; we assume, however, that even
if the transition becomes crossover, that it remains nearly 
second order \cite{rp}.
That is, we assume that Polyakov loops become light
for some range of temperatures, which we define as $T_c$.  This is
crucial to our analysis, because when $\ell$ becomes light, 
it oscillates with large amplitude about its minimum, 
and drives particle production.

(As an aside, we note that 
the free energy need not be small in the low temperature phase of a gauge 
theory.  Consider, for example, a generalized large $N$ limit, in
which $N_f/N$ is held fixed as $N_f$ and $N \rightarrow \infty$.
This large $N$ limit is ``quarky'', 
as there are $\sim N_f^2$ color singlet hadrons.  These hadrons
generate a free energy which 
is of order $N^2 \sim N N_f \sim N_f^2$ at any temperature.  
The Polyakov loop is also nonzero
at any temperature, $\ell_0 \sim N_f/N \neq 0$.
Even though $\ell_0 \neq 0$, perhaps the potential for the Polyakov
loop continues to dominate the free energy.  
However, in this case, even near $T_c$ the Polyakov loop is never light.)

For three colors, the Polyakov loop $\ell$ is complex valued, and
we take the potential
\be
{\cal V}(\ell) = 
\left( - \frac{b_2}{2} \, |\ell|^2
- \frac{b_3}{6} ( \ell^3 + (\ell^*)^3 )
+ \frac{1}{4} (|\ell|^2)^2 \right) \, b_4 \, T^4 \, ;
\label{e2}
\ee
$\ell_0(T)$ is the global minimum of ${\cal V}(\ell)$ at a
temperature $T$.  
Again, we normalize $\ell_0 \rightarrow 1$ as $T \rightarrow \infty$.

An important feature is that because the Polyakov loop is the trace of a phase
factor, it is a dimensionless field; the only scale to make up
the correct powers of dimension is the temperature.  This accounts
for the overall $T^4$ in ${\cal V}(\ell)$.  
The coefficients $b_2$, $b_3$, and $b_4$ are then fitted to
agree with Lattice data for the pressure in the pure glue
theory at $T\ge T_c$; the
pressure for $T < T_c$ is taken to vanish.
Of course the pressure in the pure glue theory is {\it not}
the same as in QCD; for instance, the ideal gas value of the pressure
changes with the addition of quarks.  Lattice simulations have
also shown that $T_c$ changes as well; 
with $2+1$ flavors one finds
$T_c \sim 180$~MeV~\cite{three,karsch3}.
Lattice simulations have demonstrated, however, that
even with dynamical quarks, 
the pressure, divided by the ideal gas value, is a nearly 
{\it universal} function of
$T/T_c$ \cite{rp,karsch3}.  We use this remarkable property in our fit,
taking $b_2(T/T_c)$, and $b_3$, from the pure glue theory.  The
overall constant $b_4$ is rescaled by the ratio of the ideal
gas terms between QCD, with three massless flavors, and the pure
glue theory.  

The pressure in the pure glue theory for $T > T_c$ \cite{karsch2}
is described with the constant values $b_3=2$ and 
$b_4\approx 0.6061$.  In QCD, we take the same value of $b_3$,
and rescale $b_4$ by the appropriate ratio of ideal gas values,
$b_4\approx 0.6061 \cdot 47.5/16$.  In the spirit of mean field theory,
we take the same values for $b_3$ and $b_4$ below $T_c$.
Below $T_c$, an approximate expression for the string tension is 
$\sigma(T)=1.21 \sqrt{\sigma_0^2- 0.99 T^2/(0.41)^2}$~\cite{karsch1},
where $\sigma_0$ is the zero temperature string tension, $=1$~GeV/fm.
The mass of the Polyakov loop is then used to fix $b_2$, 
$m_\ell^2=(\sigma(T)/T)^2\propto -b_2 b_4 T^2$. 

\begin{figure}
\begin{center}
\epsfysize=5cm
\leavevmode
\hbox{ \epsffile{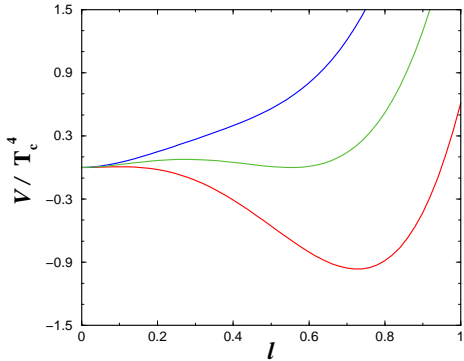}}
\end{center}
\caption{${\cal V}(\ell)/T_c^4$ for temperatures
$T/T_c=0.98$, 1, 1.02 (top to bottom), respectively.}
\label{pot}
\end{figure}
In Fig.~\ref{pot} we show ${\cal V}(\ell)$ in the vicinity of $T_c$, which is
defined as the temperature were the two local minima of ${\cal V}(\ell)$ are
degenerate. Notice that the potential
changes {\it extremely} rapidly with temperature about $T_c$.  
This is due primarily to the rapid change of the coefficient $b_2$ with
temperature.  This can be seen either above or below $T_c$.  
Above $T_c$, the ``bump'' in the trace of the energy momentum tensor,
times $1/T^4$, is due to the variation 
of $b_2$ \cite{rp}; it can also be seen from
the extremely rapid decrease of the Debye screening mass, $\sim b_2$,
from $2 T_c$ to $T_c$, by a factor of ten \cite{karsch1}.  Below
$T_c$, $b_2$ changes rapidly because of the quick rise in the string
tension as the temperature decreases \cite{karsch1}.  Physically,
this occurs because the Polyakov loop is light at $T_c$,
and very heavy at zero temperature, $m_\ell(0) \geq 1.5$~GeV.

Except near the critical temperature, the Polyakov loop is heavy,
so its fluctuations, and particle production, are small.
Only near $T_c$ does the Polyakov loop become light and generate pions.

At $T_c$, because there is a first order transition, there 
are two degenerate minima, at $\ell = 0$ and $\ell \approx 0.56$. 
As can be seen from the figure, however, the barrier between these
two minima is small.  This is because the pure glue theory
has a weakly first order transition when $N=3$.  It is so weak that
this plays no part in our analysis.  Thus whether the theory with
dynamical quarks has a first order transition, or merely a crossover,
is of little consequence; all that matters is that $\ell$ becomes light.

(The rapid variation of the potential with temperature, and the shallowness
of the barrier between the two minima at $T_c$, is in contrast to
models of the chiral transition \cite{bubble}.  For the chiral transition,
the potentials change much more slowly with temperature, and the barriers
to nucleation are significant; see, {\it e.g.}, fig. (1) of \cite{bubble}.)

The complete effective theory is described by adding a kinetic term
for $\ell$, and coupling it to a chirally symmetric field, $\phi$:
\begin{equation}
{\cal L} \; = \; {\cal L}_\phi + 
\frac{N}{g^2}
|\partial_\mu \ell|^2 T^2 - {\cal V}(\ell) - \frac{h^2}{2} {\phi}^{2} \;
|\ell|^2 T^2 .
\label{ec}
\end{equation}
$\phi$ is the usual field describing pions and the sigma meson,
with $O(4)$ global symmetry.  The model can also be extended to
three flavors, with kaons, {\it etc.}

Because the Polyakov loop 
is an effective field, normalized so that $\ell_0 \rightarrow 1$
at high temperature, the coefficient of the kinetic term is not
automatic.  
In the following analysis we require the coefficient
of the kinetic term for the Polyakov loop, $\ell$, as it varys
in time.  We do not know this, so 
in (\ref{ec}) we take the coefficient of the kinetic term
for $SU(3)$ Wilson lines, $\tilde{\ell}_a$, which vary in space \cite{rp};
we then assume a Lorentz invariant form to obtain the coefficient
for time varying fields.  While this is just a guess, taking
$\alpha = g^2/(4 \pi) = 0.3$ at $T_c$ gives
a coefficient $N/g^2 \sim 0.8$, which is about unity and so reasonable.

The coupling constant for the Polyakov loop with mesons,
$h^2$, could be fixed by knowing $\ell_0(T)$ and the meson masses
above $T_c$.  Since $\ell_0(T)$ is not known \cite{rp}, our estimate
is imprecise.  Comparing to Lattice data of Gavai and Gupta for
$m_\pi(T)$ at $T \geq T_c$ \cite{gavai},
$m_\pi/T=2\pi$ at $T\rightarrow\infty$ gives $h^2\approx 29$;
at $T/T_c=2.9$, $m_\pi/T=4.29$,
and the Hartree relation $m_\pi^2=\lambda^2(T^2/2-v^2)+h^2 T^2
|\ell|^2$ (which follows from eqs.~\ref{ec},\ref{hartree}),
yields $h^2=9$. In the actual calculations described below we employed
$h^2 = 22$ as an intermediate value.

The lagrangian for the $\phi$ field is standard:
\begin{equation}
{\cal L}_\phi \; = \;
\frac{1}{2}(\partial_\mu {\phi})^2
- \frac{\lambda^2}{4} \left( {\phi}^2 - v^2 \right)^2 + H\sigma\; .
\label{hartree}
\end{equation} 
At $T=0$ chiral symmetry is spontaneously broken, and the expectation values
are $\langle\sigma\rangle=f_\pi$, $\langle\vec{\pi}\rangle=0$. Further,
the PCAC relation gives $H=f_\pi m_\pi^2$,
were $m_\pi=138$~MeV is the pion mass in the vacuum state at $T=0$. Then 
$v^2=f_\pi^2-m_\pi^2/\lambda^2$, 
and assuming a coupling $\lambda^2=20$ yields a sigma mass in
vacuum of $m_\sigma^2=2\lambda^2f_\pi^2+m_\pi^2\approx(600$~MeV$)^2$.

There are many other terms which we could have included in the effective
lagrangian.  In particular, we really should have included terms
linear in the Polyakov loop.  In the pure glue theory, such terms
are prohibited by the global $Z(3)$ symmetry of 't Hooft, but it is
expected to occur with dynamical quarks \cite{banks}.  
A term linear in the Polyakov loop will 
change the first order transition (in the pure
glue theory) to crossover.  We have ignored terms linear in the
Polyakov loop because the free energy is small below $T_c$, so they
must be small.

In this vein, we note the results of Digal, Laermann, and Satz \cite{digal}.
They obtained Lattice results for two flavors of light,
dynamical quarks.  As expected for a second order chiral phase
transition, about $T_c$ there is a peak in the chiral susceptibility,
which becomes more pronounced as the quark mass decreases.
For the range of quark masses studied, however, 
the peak in the susceptibility for the Polyakov loop also continues
to grow.  A term linear in the Polyakov loop
should eventually cut off the divergence in the susceptibility for the
Polyakov loop; thus their results also suggest that in the effective
lagrangian, the effects of
all terms linear in $\ell$ (times one, $\phi^2$, etc.) are small.  

To compute particle production we take a simple
dynamical picture.  We assume an instantaneous quench from an initial
temperature $T_+ > T_c$, to a temperature $T_- < T_c$ \cite{dcc2}.
We stress that because the 
$\ell$-potential varies so rapidly with temperature around $T_c$,
the assumption of a quench appears natural.  We then evolve
the field, given its value for the minimum at $T_+$, with the
potential at $T_-$.  The $\ell$-field is no longer at a minimum
at $T_-$, and so it rolls down the potential, and oscillates
around the new minimum at $\ell_0 = 0$.  These oscillations in
$\ell$ couple to $\phi$, and thereby produce pions and sigmas.
For the purposes of illustration we only include the production
of pions, ignoring that of sigma mesons.

We perform the standard decomposition of the pion field in terms of creation
and annihilation operators times adiabatic mode functions $\vec{\pi}_k(t)$
\cite{preheating,boyanovsky}. The equation of motion for the mode
functions is
\bea
\frac{d^2 \vec{\pi}_k(t)}{dt^2} &=& - \Omega_k^2 \vec{\pi}_k(t)\; ,\\
\Omega_k^2 &=& k^2 + m_\pi^2 + \lambda^2\langle\vec{\pi}^2(t)\rangle
+ h^2 |\ell|^2 T^2 \; .
\eea
The effective pion mass is calculated assuming Hartree type factorization,
\be
\vec{\pi}^2(t,\vec{x}) \simeq \langle\vec{\pi}^2(t)\rangle
\ee
which requires a subtraction at $t=0$ \cite{boyanovsky}
\bea
\langle\vec{\pi}^2(t)\rangle &=& 
\int \frac{d^3k}{(2\pi)^3} |\vec{\pi}_k|^2\nonumber\\
 & &\hspace*{-1.2cm}
 =d_\pi \int \frac{d^3k}{(2\pi)^3} \left(\frac{N_k(t)}{\Omega_k(t)}
+\frac{1}{2\Omega_k(t)} -
\frac{1}{2\Omega_k(t=0)}\right)\; .
\eea
$d_\pi=3$ counts the number of isospin states.
Inserting a cutoff $\Lambda$, which is assumed to be larger than all other
mass scales, gives
\bea
\langle\vec{\pi}^2(t)\rangle &\simeq &
\frac{d_\pi}{8}\left[ c_0^2\log\frac{4\Lambda^2}{c_0^2}-
c_t^2\log\frac{4\Lambda^2}{c_t^2} \right] \nonumber\\
&+&
d_\pi\int \frac{d^3k}{(2\pi)^3} \frac{{\cal N}_k}{\Omega_k} \; ,
 \label{pifluc}
\eea
where $c_0^2 = m_\pi^2 + h^2 |\ell(t=0)|^2 T^2 $ and $c_t^2=
m_\pi^2 + \lambda^2\langle\vec{\pi}^2(t)\rangle
+ h^2 |\ell(t)|^2 T^2 $. 
In the above equations, ${\cal N}_k$ denotes the occupation number of 
a mode with given isospin and with momentum $k$,
subtracted for vacuum fluctuations,
\be
{\cal N}_k = \frac{\Omega_k}{2} \left(\frac{|\dot{{\pi}}_k|^2}{\Omega_k^2} 
+ |{\pi}_k|^2 \right) - \frac{1}{2} \; .
\ee
As initial conditions we consider the case that the state at $t=0$ contains
vacuum fluctuations only, 
such that $\vec{\pi}_k(t=0)=\sqrt{1/2\Omega_k(t=0)}$,
$d\vec{\pi}_k(t=0)/dt=-i\sqrt{\Omega_k(t=0)/2}$, and ${\cal N}_k(t=0)=0$. 

Similarly, in the classical equation of motion for the 
Polyakov loop,
we include the backreaction from the produced pions in the Hartree
approximation, replacing $\vec{\pi}^2(t,\vec{x})$ by
$\langle\vec{\pi}^2(t)\rangle$ as given in~(\ref{pifluc}).
We neglect fluctuations of $\ell$ here, which should be a reasonable
approximation as long as the energy of the produced pions is well below 
that for the Polyakov loop at $t=0$. Of course, a more refined 
treatment is necessary in order to trace the time evolution to the
point where the coherent oscillation of $\ell$ has dissipated its
entire energy into pions, since produced mesons will scatter off the
Polyakov loop condensate and cause it to decohere.

We performed numerical simulations for the following values of the
parameters: $\Lambda=10f_\pi\sim1$~GeV, $T_+=1.02$, $T_-=0.975$. In
preheating after inflation~\cite{preheating}, the coupling constant is
very small, $\sim10^{-12}$, and the inflaton field
oscillates many times.  Particle production occurs continuously
during these oscillations, and resonance bands develop for modes
in phase with the driving inflaton field.  
In the present model, both the coupling of the Polyakov loop to pions,
$h^2 = 22$, as well as the self coupling of the 
pions, $\lambda^2=20$, are large. Consequently pion production
happens rather quickly, within a few oscillations.

\begin{figure}
\begin{center}
\epsfysize=5cm
\leavevmode
\hbox{ \epsffile{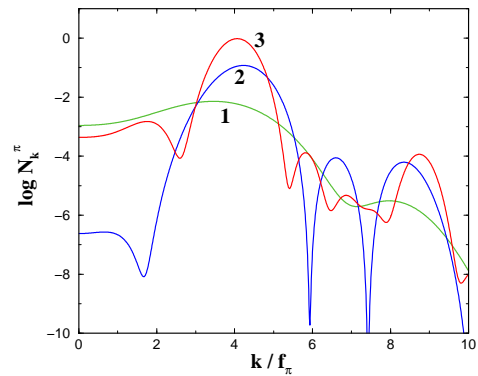}}
\end{center}
\caption{Occupation numbers of the pion modes at times $t=4n/3f_\pi$,
for $n=1..3$.}
\label{occ}
\end{figure}
In fig.~\ref{occ}
we show the logarithm of the produced pion occupation number versus
momentum.  The distribution exhibits less pronounced resonance bands
than in the case where the contribution of fluctutations to $\Omega_k$ is
neglected (compare also fig.~4 and fig.~5 in~\cite{ad_ove}). 
That is because $\lambda^2\langle\vec{\pi}^2(t)\rangle$ is
time dependent, and shifts the resonance bands to other $k$ modes.

\begin{figure}
\begin{center}
\epsfysize=5cm
\leavevmode
\hbox{ \epsffile{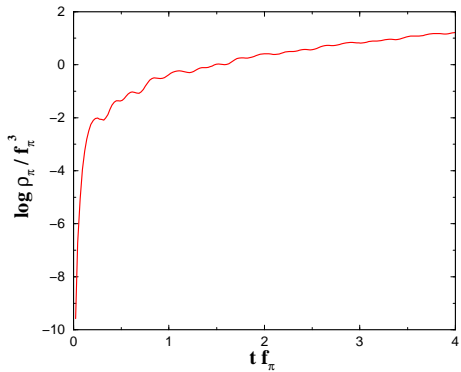}}
\end{center}
\caption{Pion density as a function of time.}
\label{totpi}
\end{figure}
Integrating over momentum, 
\be
\rho_\pi = \frac{d_\pi}{2\pi^2} \int_0^\Lambda dk \, k^2 {\cal N}_k \;
\ee
we obtain the total density of produced pions.
As shown in fig.~\ref{totpi},
the pion density increases very rapidly:
Parametric resonance lets the pion
density increase exponentially over the depicted time interval.

The total pion density at time $t=4/f_\pi$ after the quench
is approximately $\rho_\pi= 0.35/$fm$^3$. From $\rho_\pi$, the 
total number of pions per unit rapidity, $dN/dy$, is estimated as follows.
Longitudinal length in the beam direction is a scale factor,
$a$, times length in rapidity.
For one dimensional Bjorken expansion~\cite{bjorken},
$a = \tau$, where $\tau$ is the proper time.
If the expansion is isentropic,
and $s$ is the entropy density, then $s a$ is a constant;
taking $s \sim T^3$, a rough estimate of $a(T_c)$ at RHIC
is $\sim 10$~fm.  Then $dN/dy = \rho_\pi \pi R^2 a(T_c)$.
For a nucleus with $A \sim 200$, even without
transverse expansion $R\simeq 7$~fm, so that 
$dN/dy \approx 1300\rho_\pi$~fm$^3$ $\approx 460$.
If transverse expansion increases $R$ by $50\%$,
$dN/dy$ doubles and agrees roughly with data for central
collisions at BNL-RHIC~\cite{phobos}.
\begin{figure}
\begin{center}
\epsfysize=5cm
\leavevmode
\hbox{ \epsffile{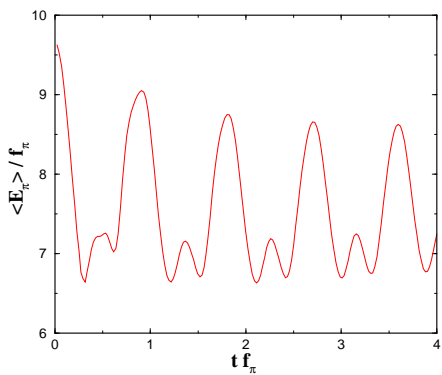}}
\end{center}
\caption{Mean energy per pion (in a given event and domain)
as a function of time.}
\label{ene_pi}
\end{figure}
Fig.~\ref{ene_pi} shows the average energy per produced pion as a function
of time.  Unlike a thermal bath, the energy of {\it all} pions oscillates
in {\it phase} with the background $\ell$-field, through
the terms $\langle\vec{\pi}^2\rangle$ and $h^2 |\ell|^2$.
Because there is a lot of energy in the $\ell$-field, the
pion energy is relatively large, and there
are large fluctuations about the average.  From Fig.~\ref{ene_pi},
the time-averaged 
root-mean-square (RMS) fluctuation of the average pion
momentum is $\sqrt{\langle p^2\rangle-\langle p\rangle^2}
/\langle p\rangle\approx10.3\%$.
We have taken the ensemble average, i.e.\ the average over events, to be given
by the time average from $t=0$ up to $t=4/f_\pi$.

This is our principal result.  In a single domain, 
as for figs.~\ref{occ} and~\ref{ene_pi}, decay of
a classical Polyakov loop condensate generates large fluctuations.
In this picture, pions are produced in a pulse near $T_c$,
from the ``ringing'' of the Wilson line into pions.

There are two effects which act to reduce the effect dramatically.
First of all, experimentally it is necessary to average over
many domains.  The size of each domain is determined by how
the quench ends, and involves both
quantum mechanical and semi-classical processes.
A lower bound for the size of a single
domain is presumably given by the correlation length of the $\ell$-field
at $T_c$, $\xi \sim 1/m_\ell(T_c)$.  From our estimates above,
because $\ell$ becomes light at $T_c$, this correlation length
is relatively large, $\xi \sim 1$~fm. Assuming one samples experimentally
the full transverse size of the nucleus, and over one unit of rapidity,
gives $N_D=\pi R^2 a(T_c)/ (4\pi \xi^3/3)\sim 300$ domains.
While perhaps semi-classical methods are suspect when each domain
gives $\sim 1.5$ pions, if we average 
over $300$ domains, the RMS fluctuation of the average momentum per pion
is reduced to $\sim 10.3\%/\sqrt{N_D}=0.6\%$. 
Averaging over many domains also acts to smooth out the 
momentum distribution of fig.~\ref{occ}, to leave a distribution
which is closer to exponential.  

While fluctuations $\sim 0.6\%$ are small, they are
experimentally measurable. For central collisions at the CERN-SPS,
$\surd{s}=18A$~GeV, the NA49 experiment
compared the following two
distributions of the average pion momentum~\cite{NA49,rss,heiselberg}.
The first was the
event-by-event distribution of the actual data, the variance of which
reflects intrinsic physical correlations in the event. The second
is the same distribution in  mixed events, constructed by picking
particles at random from different events. The variance of mixed events is
determined by purely statistical fluctuations, such as those due
to finite particle number. To within $0.05\%$, however, the RMS fluctuations
of the two distributions are the same. 

Apparently, our model does not apply at SPS energies.
One possibility is that the model is wrong.  
Another is that 
too large a region in rapidity, {\it etc.} was averaged over.
It would be interesting to bin the data in increasingly small
bins in rapidity, until one is limited by the usual statistical
uncertainty, $\sim 1/\sqrt{N_{\Delta y}}$, where 
$N_{\Delta y} = (dN/dy) \Delta y$
is the number of particles in a rapidity bin of width $\Delta y$.

It is also possible that
fluctuations at the SPS are washed out by scattering in a hadronic phase.
We assume that particles produced at $T_c$ essentially flow to
the detectors without further collisions, which is of course an
idealized scenario.
Collisions in a hadronic phase act to erase intrinsic fluctuations
generated at $T_c$, and leave purely statistical fluctuations.  
We leave a detailed investigation of this point for future work. 
It is not clear, however, how easy it is for hadronic scattering
to wash out fluctuations at $T_c$, due to expansion inherent
in a heavy ion collision.  Pions are
nearly Goldstone bosons, whose scattering is suppressed at low energy by
powers of momentum.  For example, 
in model calculations \cite{ad_ove}, even though the
self coupling of pions is large, 
pions can scatter into the zero mode of the pion field more
rapidly than they can scatter out.

Our estimate of $dN/dy$ is just to tune the parameters of the model,
but fluctuations in mean momentum $\sim 0.6 \%$
should be a qualitatively correct prediction.
Indeed, one can easily imagine how the fluctuations could even be larger.  
If the average size of the domains, $\xi$, 
is not $\sim 1$ but $\sim2$~fm, then in one unit of rapidity 
there are only 40 domains instead of 300.
Each domain generates about ten pions, and 
the RMS fluctuations of the average momentum per pion increases to
$\sim 10.3\%/\sqrt{40}=1.7\%$. Notice, too, that 
as particle production increases
exponentially in time, the distribution is dominated by those domains
which last for the longest time.

If indeed hadron production at the confinement transition is dominated
by the decay of a Polyakov loop condensate,
we also expect nonstatistical (intrinsic) fluctuations
of the pion multiplicity from event to event.
For example, the RMS fluctuation of the multiplicity during the last
1~fm/c depicted in fig.~\ref{totpi} is $\sim24\%$ 
in a single domain; for 300 domains, this gives fluctuations
$\sim 1.4\%$. 

Recently, fluctuations of the electric charge were
analyzed~\cite{charged}. 
Based upon a quasi-particle picture, it is argued that
that RMS fluctuations of the charge,
measured in a rapidity window $\Delta  y\sim1$, 
should be much smaller in a quark-gluon plasma than in a pion gas.
Thus, if a thermal and locally neutral plasma were produced,
and the pions emerged from the hadronization of that plasma
with small relative rapidity, the charge
fluctuations in the final state remain small: given a charged
pion in $\Delta  y$, the probability for its partner with
opposite charge to have rapidity within $\Delta  y$ is large.
In the present model, on the other hand, pions produced from
oscillations of the electrically neutral Polyakov loop
are spread over a rapidity interval of $\sim\pm2$ (see
fig.~\ref{occ}), and the number of $\pi^+$ and $\pi^-$
in a given rapidity window $\Delta  y\sim1$ fluctuates
independently. Thus, the RMS fluctuations
of the charge should be large, i.e.\
like that of hadrons, not quarks.

The fluctuations are dominated by the behavior of the system
as it goes through $T_c$.  
To first approximation, it does not matter how hot the system is
initially.  Consequently, the magnitude of the fluctuations should
be approximately similar in collisions at RHIC and at the LHC.
The only change with increasing the energy
of the colliding nuclei is a change in the scale factor $a(T_c)$,
which produces a higher multiplicity at higher energy.  
As a constant number of pions is produced per domain, the width in
rapidity should be chosen so that there is the same number of particles
per rapidity bin.
Higher energies can help, albeit indirectly: 
transverse expansion may be larger at higher energies, which
prevents rescattering in the hadronic phase
from washing out fluctuations induced at $T_c$.

We have made numerous approximations in the present analysis.  
A more careful study is certainly possible.  The
potential for the Polyakov loop in QCD, and the coupling
to pions, can be computed from numerical simulations on the lattice,
both without and with dynamical quarks.
Many other improvements can also be made:
introducing realistic models for the space-time
dependence of the temperature, the effect of expansion
in the equations of motion for $\ell$ and $\phi$, {\it etc.}

We have not attempted a better analysis here because STAR
and other detectors will soon decide if heavy 
ion collisions at RHIC do or do not
display intrinsic fluctuations on top of
purely statistical fluctuations.  We eagerly await this analysis.

{\it Note added:}  After this paper was submitted for publication,
many results were announced at Quark Matter 2001.
CERES reported data from the
SPS indicating fluctuations in the average transverse pion
momentum $\sim 3 \%$ \cite{appel}.  At 
RHIC, STAR found that the same fluctuations appear to be
much larger, $\sim 8 \%$ 
\cite{flucs}.  STAR also reported that pion interferometry
appears to indicate ``explosive'' behavior \cite{flucs,explosive}.
We have recently suggested how the Polyakov loop model
might generate explosive particle production and large fluctuations \cite{qm}.

\acknowledgements
A.D.\ acknowledges support from a DOE Research Grant, 
Contract No.\ DE-FG-02-93ER-40764; R.D.P.\ from
DOE grant DE-AC-02-98CH-10886.  R.D.P.\ thanks 
the Heavy Ion Group at Wayne State University for discussions,
including R.\ Bellwied, S.\ Gavin, C.\ Pruneau, and S.\ Voloshin.

\end{document}